# Measurement of the response of a Ga solar neutrino experiment to neutrinos from an $^{37}$Ar source

J. N. Abdurashitov, V. N. Gavrin, S. V. Girin, V. V. Gorbachev, P. P. Gurkina,
T. V. Ibragimova, A. V. Kalikhov, N. G. Khairnasov, T. V. Knodel, V. A. Matveev,
I. N. Mirmov, A. A. Shikhin, E. P. Veretenkin, V. M. Vermul, V. E. Yants, and G. T. Zatsepin
*Institute for Nuclear Research of the Russian Academy of Sciences, Moscow 117312, Russia*

T. J. Bowles, S. R. Elliott, and W. A. Teasdale
*Los Alamos National Laboratory, Los Alamos, NM 87545 USA*

B. T. Cleveland, W. C. Haxton, and J. F. Wilkerson
*Department of Physics, University of Washington, Seattle, WA 98195 USA*

J. S. Nico
*National Institute of Standards and Technology, Gaithersburg, MD 20899 USA*

A. Suzuki
*Research Center for Neutrino Science, Tohoku University, Aramaki, Aoba, Sendai, Japan*

K. Lande
*Department of Physics and Astronomy, University of Pennsylvania, Philadelphia, PA 19104 USA*

Yu. S. Khomyakov, V. M. Poplavsky, and V. V. Popov
*Institute of Physics and Power Engineering, Obninsk 249020, Kaluga region, Russia*

O. V. Mishin, A. N. Petrov, B. A. Vasiliev, and S. A. Voronov
*OKB Mechanical Engineering, Nizhny Novgorod 603074, Russia*

A. I. Karpenko, V. V. Maltsev, N. N. Oshkanov, and A. M. Tuchkov
*Beloyarsk Nuclear Power Plant, Zarechny 624250, Sverdlovsk region, Russia*

V. I. Barsanov, A. A. Janelidze, A. V. Korenkova, N. A. Kotelnikov,
S. Yu. Markov, V. V. Selin, Z. N. Shakirov, A. A. Zamyatina, and S. B. Zlokazov
*Institute of Nuclear Materials, Zarechny 624250, Sverdlovsk region, Russia and*

An intense source of $^{37}$Ar was produced by the $(n, \alpha)$ reaction on $^{40}$Ca by irradiating 330 kg of calcium oxide in the fast neutron breeder reactor at Zarechny, Russia. The $^{37}$Ar was released from the solid target by dissolution in acid, collected from this solution, purified, sealed into a small source, and brought to the Baksan Neutrino Observatory where it was used to irradiate 13 tonnes of gallium metal in the Russian-American gallium solar neutrino experiment SAGE. Ten exposures of the gallium to the source, whose initial strength was 409 ± 2 kCi, were carried out during the period April to September 2004. The $^{71}$Ge produced by the reaction $^{71}$Ga$(\nu_e, e^-)^{71}$Ge was extracted, purified, and counted. The measured production rate was $11.0^{+1.0}_{-0.9}$ (stat) ± 0.6 (syst) atoms of $^{71}$Ge/d, which is $0.79^{+0.09}_{-0.10}$ of the theoretically calculated production rate. When all neutrino source experiments with gallium are considered together, there is an indication the theoretical cross section has been overestimated.

## I. INTRODUCTION

To verify that their efficiencies are well understood, the two gallium solar neutrino experiments have measured their response to reactor-produced neutrino sources of known activity. Sources of $^{51}$Cr were used in these experiments, as reported in [1] for SAGE and in [2] for GALLEX/GNO. Table I gives some of the relevant details of these experiments and on the experiment with an $^{37}$Ar source that is the subject of this article.

There are several advantages of an $^{37}$Ar source compared to a $^{51}$Cr source, which we enumerate here. A major advantage is that the desired active isotope must be chemically separated from the target following irradiation. Although this involves an additional processing step, this separation serves to remove almost all impurities that are present in the target. Thus an $^{37}$Ar source, in contrast to a $^{51}$Cr source, which is made by irradiating Cr metal, without any subsequent purification, can be made practically free of radioactive impurities. Other advantages of $^{37}$Ar compared to $^{51}$Cr are that the half-life is longer (35 d compared to 27 d), thus giving more time to prepare the source and to make measurements, that the neutrino energy is greater (811 keV compared to 747 keV), thus giving a higher cross section, that the decay is purely to the ground state (100% compared to 90%), thus giving a monoenergetic neutrino source, and that there are no accompanying $\gamma$ rays (except for inner bremsstrahlung [4]), thus requiring little shielding and yielding a very compact source. Finally, since nearly 97% of Ca is $^{40}$Ca, no isotopic separation is required before irradiation. In contrast, to make a $^{51}$Cr source, the irradiated isotope $^{50}$Cr must be enriched as its content in natural Cr is only 4.3%.

TABLE I: Comparison of source experiments with Ga. When two uncertainties are given, the first is statistical and the second is systematic. When one uncertainty is given, statistical and systematic uncertainties have been combined in quadrature. The values of $R$ for GALLEX are from [3].

| Item | GALLEX Cr1 [2, 3] | GALLEX Cr2 [2, 3] | SAGE $^{51}$Cr [1] | SAGE $^{37}$Ar |
|---|---|---|---|---|
| Source production | | | | |
| Mass of reactor target (kg) | 35.5 | 35.6 | 0.512 | 330 |
| Target isotopic purity | 38.6% $^{50}$Cr | 38.6% $^{50}$Cr | 92.4% $^{50}$Cr | 96.94% $^{40}$Ca (natural Ca) |
| Source activity (kCi) | $1714^{+30}_{-43}$ | $1868^{+89}_{-57}$ | $516.6 \pm 6.0$ | $409 \pm 2$ |
| Specific activity (kCi/g) | 0.048 | 0.052 | 1.01 | 92.7 |
| Gallium exposure | | | | |
| Gallium mass (tonnes) | 30.4 (GaCl$_3$:HCl) | 30.4 (GaCl$_3$:HCl) | 13.1 (Ga metal) | 13.1 (Ga metal) |
| Gallium density ($10^{21}$ $^{71}$Ga/cm$^3$) | 1.946 | 1.946 | 21.001 | 21.001 |
| Measured production rate $p$ ($^{71}$Ge/d) | $11.9 \pm 1.1 \pm 0.7$ | $10.7 \pm 1.2 \pm 0.7$ | $14.0 \pm 1.5 \pm 0.8$ | $11.0^{+1.0}_{-0.9} \pm 0.6$ |
| $R = p(\text{measured})/p(\text{predicted})$ | $1.00^{+0.11}_{-0.10}$ | $0.81^{+0.10}_{-0.10}$ | $0.95 \pm 0.12$ | $0.79^{+0.09}_{-0.10}$ |

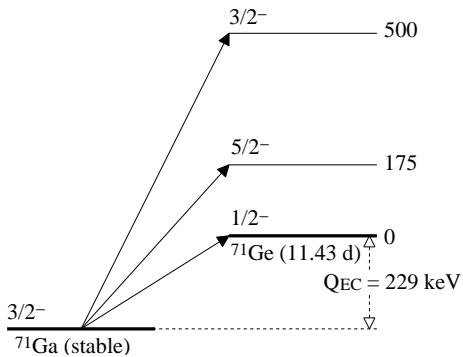

FIG. 1: Levels in $^{71}$Ge that can be reached by capture in $^{71}$Ga of a neutrino from $^{37}$Ar or $^{51}$Cr sources.

The use of $^{37}$Ar to measure the response of radiochemical solar neutrino detectors was originally proposed by Haxton [5], who pointed out that the neutrino energy from $^{37}$Ar decay is very near that of the principal neutrino line produced by $^7$Be electron capture in the Sun at 863 keV. A practical method to make an intense $^{37}$Ar source by the $(n, \alpha)$ capture reaction on $^{40}$Ca at a reactor with a high flux of fast neutrons was given by Gavrin *et al.* [6].

Both $^{37}$Ar and $^{51}$Cr can excite only the lowest three energy levels in $^{71}$Ge, as shown in Figure 1. The cross section for the transition to the ground state of $^{71}$Ge is well determined by the decay constant; the strength of the transitions to the two excited states, at 175 keV and 500 keV, is much more poorly known, but can be estimated from $(p, n)$ scattering [7] measurements. As evaluated by Bahcall [8], 95% of the cross section for the neutrinos from $^{37}$Ar is to the ground state, the state at 500 keV contributes 3.6% to the cross section, and the state at 175 keV contributes the remainder, although only an approximate upper limit for its strength has been measured.

TABLE II: Decay modes of $^{37}$Ar and the energy released. The atomic-electron binding energies are from [9]. The branching fractions are based on the ratios $L/K = 0.0987 \pm 0.003$ [10] and $M/L = 0.104^{+0.006}_{-0.003}$ [11]. The internal bremsstrahlung fraction and average energy are estimated for 1s-electron capture using Eq. (2) and (3), respectively, in [12], and for 2s-electron and p-electron capture are taken from Fig. 1 of [13]. These calculations have been verified experimentally for $^{37}$Ar in [13] (spectrum) and [14] (absolute intensity). See also [15, 16].

| Decay mode | Atomic energy release (keV) | Fraction of $^{37}$Ar decays | Energy per $^{37}$Ar decay (keV) |
|---|---|---|---|
| $K$ capture | 2.8224 | $0.9017 \pm 0.0024$ | $2.5450 \pm 0.0068$ |
| $L$ capture | 0.2702 | $0.0890 \pm 0.0027$ | $0.0240 \pm 0.0007$ |
| $M$ capture | 0.0175 | $0.0093^{+0.0006}_{-0.0004}$ | 0.0002 |
| 1s int. brems. | 325 (average) | $\sim 0.0005$ | $\sim 0.16 \pm 0.02$ |
| 2s int. brems. | 325 (average) | $\sim 0.00007$ | $\sim 0.021 \pm 0.002$ |
| p int. brems. | $\sim 10$ (average) | $\sim 0.00007$ | $\sim 0.0007$ |
| Total | | | $2.751 \pm 0.021$ |

## II. DECAY OF $^{37}$Ar

$^{37}$Ar decays to $^{37}$Cl with a half-life of $35.04 \pm 0.04$ d [17], as shown in Fig. 2. The decay is solely by electron capture and the $Q$-value is 813.5 keV.

The decay energy that heats the source is given in Table II. The major uncertainty in the source heating is from the fraction of internal bremsstrahlung (IB) decays and is estimated to be $\sim 10\%$. A Monte Carlo calculation showed that the IB $\gamma$ rays not absorbed in the shield outside the source take away an inconsequential $0.2 \pm 0.1\%$ of the source energy. We thus assume the heat deposited in the source following IB is simply the average decay energy.

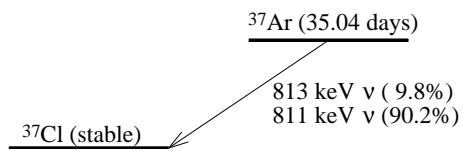

FIG. 2: $^{37}$Ar decay scheme showing the neutrino energies.

### III. SOURCE PRODUCTION

The source was made by irradiating calcium oxide in the fast neutron breeder reactor BN-600 at Zarechny, Russia. The total fast flux at this reactor is $2.3 \times 10^{15}$ neutrons/(cm$^2$ s), of which $1.7 \times 10^{14}$ neutrons/(cm$^2$ s) have energy above the 2-MeV threshold of the production reaction $^{40}$Ca$(n, \alpha)^{37}$Ar. Nineteen irradiation assemblies, each of which contained 17.3 kg of CaO (12.36 kg Ca), were placed in the blanket zone of the reactor. Irradiation began on 31 October 2003 and continued until 12 April 2004, the normal reactor operating cycle. After a cooling period of a week, the assemblies were removed from the reactor and moved to a hot cell, where each was opened and from which the capsule containing the CaO was removed. These capsules were transported to an extraction facility at the Institute of Nuclear Materials, where each capsule was cut open in a vacuum system and the CaO dissolved in a nitric acid solution.

$^{37}$Ar was extracted from this acid solution by a He purge, purified, and then stored on charcoal at LN$_2$ temperature. The purification involved flowing the gas over zeolite at room temperature, followed by two Ti absorbers, operating at 900–950 °C and 400–450 °C. The purified $^{37}$Ar, whose volume was ~2.5 l at STP, was then adsorbed on another charcoal trap and measurements of gas volume and isotopic composition were made.

As the last steps of source fabrication, the purified Ar was transferred to a pre-weighed source holder, which consisted of a sealable stainless steel vessel with a volume of ~180 ml. Inside this vessel was 40 g of activated charcoal onto which the purified $^{37}$Ar was cryopumped. When essentially all the $^{37}$Ar had been adsorbed, the vessel was closed by compressing two separate knife-edge seals, one onto a copper gasket and another onto a lead gasket. The source holder was then weighed to determine the amount of $^{37}$Ar contained within. The calculated gas pressure in the source holder was ~17 atm at room temperature. To complete the source, the source holder was placed within two concentric stainless steel vessels with a Pb shield between them. These two vessels were welded shut and the heat output of the finished source was measured with a calorimeter. These procedures were completed on 29 April 2004 and the source was immediately flown by chartered plane to the Mineralnye Vody airport, close to the experimental facility at the Baksan Neutrino Observatory in the northern Caucasus mountains.

### IV. USE OF THE SOURCE AT BAKSAN

The experimental procedures and equipment were basically the same as used with the SAGE Cr experiment in 1995. These were described in detail in a previous article [1], and will only be briefly summarized here.

A diagram of the experimental area is given in Fig. 3. The gallium is contained in seven chemical reactors, designated on this figure as numbers 2, 3, 4 and 7–10, all of which are fitted with stirrers and ancillary extraction apparatus. Reactors 2–4 contained 22 tons of gallium and were not used in the experiment reported here; rather, they were used at the time of the Ar experiment for solar neutrino measurements which will be reported elsewhere.

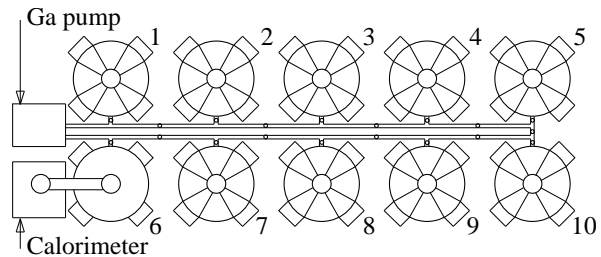

FIG. 3: Plan view of the laboratory showing the ten chemical reactors, irradiation reactor 6 with the adjacent calorimeter, and the pump for transferring Ga between reactors.

Reactor 6 has the extraction equipment removed and replaced by a Zr re-entrant tube on its axis that extends to the reactor center. A specially designed remote handling apparatus is able to grasp the $^{37}$Ar source and place it either at the center of reactor 6 or in an adjacent calorimeter. Reactors 7–10 all contain about 6.5 tonnes of gallium. To make a source exposure the gallium from two of these reactors (either 7 and 8 or 9 and 10) is extracted so as to remove any Ge that may be present and pumped into reactor 6. The source is then placed at the center of reactor 6 and the exposure begins. The first two exposures were for seven days, followed by exposures of approximately two weeks duration. At the end of exposure, the source is moved to the calorimeter and the Ga is pumped back to the original two reactors where the Ge is extracted. At the same time, a clean-up extraction of the Ga from the other reactor pair is made and its Ga is transferred to irradiation reactor 6 to begin the next exposure.

When the source arrived, it was removed from the shipping cask, a brief measurement of its $\gamma$ activity was made, and it was immediately placed into irradiation reactor 6, which had been previously filled with 13.1 tonnes of Ga metal. The first irradiation of Ga began at 04:00 on 30 April 2004 (local time, GMT+3 h), which we call the reference time.

Ten extractions were made, designated Ar 1 through Ar10. The data for each extraction are given in Table III. As a check on the extraction process, a second extraction was made after exposure 3, designated Ar 3-2. While the irradiation reactor was being emptied and then refilled, the source was moved to the calorimeter where its heat output was measured, as described in the next section.

An outline of the extraction procedure is given in [18]. A slight variation of this procedure was used for the Ar experiment: the usual volume of extraction reagents was divided into two parts. The first part was added, mixed, and removed, and then this process was repeated with the second part. This procedure of making 'two half extractions' results in reduced heating of the Ga metal, and thus a higher overall efficiency when the two extracts are combined. It was introduced with the solar neutrino extractions beginning in 1998 and has been used ever since.

The efficiency of extraction was measured by adding to the

TABLE III: Extraction schedule and related parameters. The times of exposure are given in days of year 2004.

| Extraction name | Extraction date (2004) | Source exposure Begin | Source exposure End | Source activity (kCi) Begin | Source activity (kCi) End | Solar neutrino exposure Begin | Solar neutrino exposure End | Mass Ga (tonnes) | Carrier mass ($\mu$g) | Extraction efficiency from Ga | Extraction efficiency into GeH$_4$ |
|---|---|---|---|---|---|---|---|---|---|---|---|
| Ar 1 | 6 May | 121.17 | 127.71 | 409 | 359 | 118.48 | 127.78 | 13.085 | 0 | 0.93 | 0.59 |
| Ar 2 | 14 May | 128.42 | 135.71 | 354 | 307 | 125.38 | 135.78 | 13.084 | 215 | 0.96 | 0.93 |
| Ar 3 | 29 May | 136.42 | 150.71 | 302 | 228 | 133.51 | 150.81 | 13.063 | 211 | 0.93 | 0.90 |
| Ar 3-2 | 30 May | 136.42 | 150.71 | 302 | 228 | 150.91 | 151.81 | 13.049 | 274 | 0.93 | 0.87 |
| Ar 4 | 13 Jun | 151.42 | 165.71 | 225 | 169 | 147.47 | 165.77 | 13.055 | 208 | 0.97 | 0.92 |
| Ar 5 | 28 Jun | 166.40 | 180.71 | 167 | 126 | 162.47 | 180.77 | 13.018 | 210 | 0.98 | 0.97 |
| Ar 6 | 13 Jul | 181.42 | 195.71 | 124 | 94 | 173.57 | 195.77 | 13.025 | 219 | 0.98 | 0.97 |
| Ar 7 | 28 Jul | 196.42 | 210.71 | 92 | 70 | 193.49 | 210.79 | 12.974 | 215 | 0.98 | 0.97 |
| Ar 8 | 12 Aug | 211.42 | 225.71 | 69 | 52 | 208.48 | 225.78 | 12.997 | 209 | 0.98 | 0.96 |
| Ar 9 | 27 Aug | 226.42 | 240.71 | 51 | 38 | 223.47 | 240.77 | 12.945 | 214 | 0.98 | 0.96 |
| Ar10 | 11 Sep | 241.42 | 255.71 | 38 | 29 | 238.38 | 255.78 | 12.969 | 211 | 0.98 | 0.96 |

Ga a known mass of inactive Ge carrier before the start of exposure to the neutrino source and measuring the mass of extracted Ge. The mass of added carrier is given in Table III, as is the efficiency of extraction from the metal and the overall efficiency including synthesis into the counting gas GeH$_4$.

The carrier consists of slugs of an alloy of Ge in Ga metal. Two slugs were added to the gallium in each extraction reactor (numbers 7 and 8 or 9 and 10, depending on which pair of reactors was being used) before transfer to irradiation reactor 6. The mass of each slug is known and the average concentration of Ge is determined by extraction from a large number of slugs.

Some anomalies occurred in the first extraction Ar 1. First, due to a miscommunication, carrier was not added at the start of exposure. The extraction efficiency thus cannot be determined in the usual way and it is taken to be the same as the extraction efficiency from this reactor pair in the next extraction. Second, an error was made during GeH$_4$ synthesis which resulted in the loss of some of the sample. Since inactive Ge was added to the solution before synthesis the amount lost could be determined accurately.

As noted above, a clean-up extraction was made from each new batch of gallium shortly before exposure to the source commenced. This second extraction served to remove almost all traces of Ge carrier that remained because of the inefficiency of the first extraction and also removed any $^{71}$Ge produced by solar neutrinos during the ∼2-week interval since the last extraction from this batch of Ga. Since there is no leftover carrier, the determination of the extraction efficiency is simplified, and since there is no remaining $^{71}$Ge, the analysis can assume no carryover of $^{71}$Ge from one exposure to the next.

## V. MEASUREMENT OF SOURCE ACTIVITY

Three different methods were used to measure the source activity as it was fabricated, a series of eleven measurements were made while the source was used at Baksan, and two different measurement methods were used after the source was returned to the fabrication facility. In this section we describe these methods and give the results.

### A. Measurements at Zarechny during source fabrication

In the first method, carried out after argon purification, and while the gas was put into the source holder, the total volume of gas and its isotopic composition were measured. The composition was determined with a mass spectrometer and the results are given in Table IV. The gas volume was measured by warming the charcoal trap onto which the gas had been frozen so that the gas expanded into a calibrated volume, and reading the pressure. The gas was then frozen into the source holder, and the difference in pressure between before filling and after filling implied the volume of gas in the holder was $2.665 \pm 0.048$ l at STP. Combining this with the isotopic composition and correcting for decay between the time of volume measurement and the reference time gives a source activity of $409 \pm 5$ kCi at 04:00 on 30 April 2004. The stated uncertainty has 68% confidence and includes all known systematics.

TABLE IV: Gas content of the $^{37}$Ar source 47.5 h prior to the reference time in percent by volume. The uncertainty shown is statistical; there are additional systematic components whose sum is no more than 0.8%.

| H$_2$ | $^{37}$Ar | $^{38}$Ar | $^{39}$Ar | $^{40}$Ar |
|---|---|---|---|---|
| $0.26 \pm 0.07$ | $96.57 \pm 0.13$ | $1.87 \pm 0.06$ | $0.35 \pm 0.03$ | $0.95 \pm 0.03$ |

In the second method, the source holder was evacuated and weighed before filling and then weighed again after filling with the extracted gas sample. The difference in mass was $4.400 \pm 0.042$ g at the time of filling (06:25 on 28 April), from which the activity is calculated to be $412 \pm 3$ kCi at the reference time.

In the third method, the heat output of the source was measured in a heat flow calorimeter with an inner Pb $\gamma$-ray shield and an outer massive container for thermal stabilization. The heat produced by the source was conducted away by a thermopile containing 1500 thermocouples connected in series. The calorimeter was calibrated using electrical heaters of known power and the thermopile EMF over the range of (6–8) W (the expected source power) was found to have the constant value $9.81 \pm 0.02$ mV/W. After stabilization of the calorimeter with the source the average measured thermopile

EMF was 65.9±0.03 mV at 20:00 on 28 April 2004. Applying a decay factor of 0.9740 gives a power of 6.54 ± 0.04 W at our reference time. Using the conversion factor in Table II gives the source activity at this time as 401 ± 4 kCi. The uncertainty estimate includes the calibration uncertainty, the errors in the calorimeter measuring circuits, and the uncertainties in both decay energy and $^{37}$Ar half-life.

### B. Calorimetric measurement at Baksan

The same calorimeter that was used for the SAGE $^{51}$Cr source was again used at Baksan to measure the heat output from the $^{37}$Ar source. This device has been described in [1] and the reader is referred to that publication for details. The energy released per $^{37}$Ar decay is, however, less than one-tenth that per $^{51}$Cr decay and thus modifications to the calorimeter to improve its sensitivity were necessary. Four of the eight copper heat conductors between the copper cup and the surrounding copper can were removed and voltmeters more sensitive than originally provided were used to read the thermocouple EMF.

The calorimeter was calibrated using electrical heaters made from Al and Fe and with a mock-up source made from identical materials as the real source. The temperature response with the Al heater matched that of the mock-up source very well and thus the Al measurements were used for the calorimeter calibration.

The calorimeter was used to measure the source power at the end of each gallium exposure to the $^{37}$Ar source. Two measurements were made after the final extraction Ar10, thus yielding a total of 11 measurements, which are plotted in Figure 4.

If a weighted fit is made to this data with a decaying exponential whose half-life is fixed at 35.04 d the power at the reference time is 6.907 ± 0.013 W. $\chi^2$ for this fit is 11.2 with 10 degrees of freedom (DOF) (probability = 34%). As a check, the same fit was made allowing the decay constant to be a free variable, along with the power at the reference time. The resultant best fit half-life is 34.80 ± 0.20 d, in agreement with the known value. $\chi^2$/DOF = 9.8/9 for this fit.

Using the energy release given in Table II and the conversion factors $1.6022 \times 10^{-16}$ (W s)/keV and $3.7 \times 10^{10}$ decays of $^{37}$Ar/(Ci s), the inferred source activity at the reference time was 423.5 ± 0.8 kCi. The quoted uncertainty here of 0.2% is solely from the measurement errors. There are several additional systematic uncertainties that must be included in a full uncertainty estimate. These include uncertainties due to the differences in thermal properties between the source and the calibration heaters (estimated to be 1.5%), in the energy release (0.8%, see Table II), in the incomplete absorption of the inner bremsstrahlung component of the energy release (~ 0.2%), in the $^{37}$Ar half-life (0.04%), and in the capture of some of the $\gamma$ rays from the source in the outer parts of the calorimeter, thus disturbing the 'cold' junction temperature. We assign a total uncertainty for this measurement of ±2%, or ±9 kCi.

The traditional source of uncertainty in calorimetric mea-surements is the contribution to the source heat from impurities. For the $^{37}$Ar source, the only significant impurity was $^{39}$Ar, a pure $\beta$ emitter with a half-life of 269 y. The measured $^{39}$Ar concentration at the time of source fabrication was 0.35% by volume. Since each decay gives 565 keV and the volume of the source gas was nearly 2.7 l, the heat from $^{39}$Ar decay was only 1.8 mW, 0.03% of the heat from $^{37}$Ar at the reference time. The influence of impurities on the calorimetric determination of the activity is thus negligible.

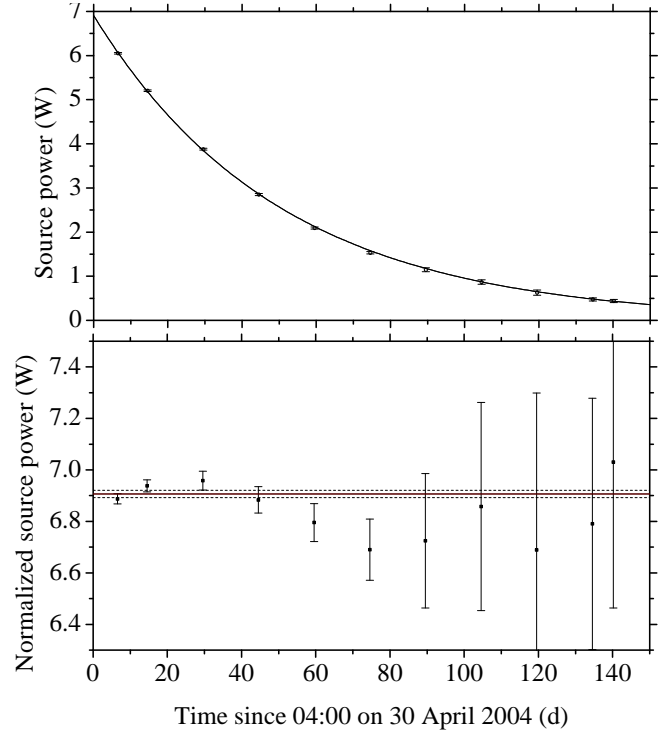

FIG. 4: Source power measurements at Baksan. The solid curve is a weighted fit of the data to a decaying exponential with the half-life of $^{37}$Ar. In the lower panel the power is normalized to the time shown and the 68% confidence band for the fit is indicated by the dashed lines.

### C. Measurement at Zarechny by $^{37}$Ar counting

The $^{37}$Ar source was returned to the fabrication facility in December 2004. The source holder was opened with a spark discharge in a vacuum system, the entire gas sample was removed, and samples of the gas were taken for activity measurement in proportional counters. At this time the $^{37}$Ar had decayed by a factor of ~300.

Because the specific activity was still very high, it was necessary to make several volume divisions to reduce the count rate to a value that was measurable in a proportional counter. These dilutions employed He as a carrier gas since the total gas volume and thus the pressure was very low. The fraction of gas transferred to the proportional counter was only about 1 part in $10^8$. To be certain that the gas content of the last sampling volume was fully transferred to the counter, spe-cial proportional counters were built which contained a side arm filled with charcoal onto which the Ar sample was cryopumped. Including the dead volume in the side arm, the counting efficiency for the $K$ peak of $^{37}$Ar was 59%. Despite the great volume division, the count rate was still very high, several hundred thousand per second. This rate was measured by continuously recording the pulses from the counter using a waveform analyzer with long time span and then counting the number of pulses during a selected time interval.

Five samples were measured in two proportional counters using different methods of volume division. Assuming an $^{37}$Ar half life of 35.04 d, the weighted average of these measurements gives a source strength at the reference time of 405.1 ± 2.7 kCi where the uncertainty includes the systematic uncertainties from counting statistics, volume division, and counting efficiency. Since the time delay from the reference time to the time of these measurements was 287 d, this result is rather sensitive to the value of the half-life used in the decay correction. The $^{37}$Ar half-life uncertainty in the most recent data compilation for this nuclear mass [17] is given as ±0.04 d, which leads to an additional uncertainty in the source strength of ±0.65%.

### D. Measurement at Zarechny by isotopic dilution

When the source was opened and gas samples were measured in proportional counters, additional samples were taken for measurement of the volume concentration of the Ar isotopes with a mass spectrometer. The gas from the source was diluted by adding a measured volume of Ar gas from the atmosphere, and the concentration of the Ar isotopes in small samples was again measured. From the combination of these measurements, using the fact that the volume of the samples taken for measurement was a small fraction of the total gas volume, we calculate the volume of $^{37}$Ar in the source at the time the initial isotopic composition measurement was made to be

$$V_{37}(\text{before}) = \frac{f_{40}V(\text{add})}{R(\text{after})D(\text{decay}) - R(\text{before})}, \quad (1)$$

where $R(\text{before})$ and $R(\text{after})$ are the measured ratios of $^{40}$Ar to $^{37}$Ar concentration by volume, before and after the dilution, respectively; $f_{40} = 0.99600$ is the fraction of $^{40}$Ar in atmospheric Ar; and $D(\text{decay})$ is the $^{37}$Ar decay factor during the time between the 'before' and 'after' measurements. The added volume of air Ar was $V(\text{add}) = 27.26 \pm 0.49$ cm$^3$ at STP.

This method was applied with several different samples taken at various times before and after dilution. As an example, for one such measurement, with the 'before' sample taken on 15 February 2005 and the 'after' sample one day later, with $R(\text{before}) = 2.70 \pm 0.02$ and $R(\text{after}) = 6.49 \pm 0.10$, the calculated volume of $^{37}$Ar in the source on 15 February 2005 by Eq. (1) was 7.707 cm$^3$ at STP, equivalent to 1.281 kCi. Extrapolating back to the reference time of 04:00 on 30 April 2004 implies an initial source strength of 408 kCi. A weighted average of all the 'before' and 'after' samples gives a source strength of 410 ± 5 kCi at our reference time. The stated uncertainty includes the error in the half-life.

### E. Summary of source strength measurements

TABLE V: Summary of all source activity measurements. The stated uncertainties include all known systematics.

| Measurement method | Activity (kCi $^{37}$Ar at 04:00 on 30 April 2004) |
|---|---|
| Volume of gas | 409.3 ± 5 |
| Mass of gas | 412.3 ± 3 |
| Calorimetry at Zarechny | 401.3 ± 4 |
| Calorimetry at Baksan | 423.5 ± 9 |
| Proportional counter | 405.1 ± 4 |
| Isotopic dilution | 410.1 ± 5 |

The results of the six activity measurements are summarized in Table V. Their weighted average is 409 ± 2 kCi. The two measurements with the calorimeter are each about 1.5$\sigma$ above and below this average, but still for all measurements $\chi^2 = 7.2$, which with 5 DOF has a probability of 21%.

## VI. $^{71}$Ge COUNTING AND EVENT SELECTION

The extracted Ge was synthesized into the counting gas GeH$_4$, mixed with inactive Xe, and inserted into special ultralow background proportional counters with a carbon-film cathode [19]. The counters were measured in an electronics system which recorded the full waveform for 800 ns after pulse onset. Both event energy and rise-time were obtained from the pulse waveform and used to select candidate $^{71}$Ge events. The system operated very stably with no need for equipment replacement during counting. The counting information is given in Table VI.

The counters were calibrated with $^{55}$Fe and Cd-Se at the start of counting and then approximately every 2 weeks until counting ended. Extreme stability was observed with the peak position rarely differing from one calibration to the next by more than 1%, the statistics on the number of calibration events. The rise-time characteristics were also highly stable.

The location of the energy acceptance windows for $^{71}$Ge events in the $L$ and $K$ peaks was set from the $^{55}$Fe calibration in the same way as for solar neutrino runs. This included our standard adjustment factors because of non-linearity in the energy scale [18].

Exactly the same procedures were used to select candidate $^{71}$Ge events as we use for solar neutrino runs. As the first step in event selection, two time cuts were applied to the data to suppress false $^{71}$Ge events produced by Rn. To reduce the effect of Rn external to the counters which may enter whenever the counting system shield is opened, a cut was made to delete the first 3 h of data after the shield is closed and counting resumes. To reduce the effect of Rn internal to the counters which may enter when the counters are filled, a second time

TABLE VI: Counting parameters. $\Delta$ is the exponentially weighted live time. The live time and $\Delta$ include all time cuts.

| Extraction name | Counter filling | | Counter name | Operating voltage (V) | Counting efficiency after rise time and energy cuts | | Day counting began in 2004 | Live time of counting (days) | $\Delta$ |
|---|---|---|---|---|---|---|---|---|---|
| | GeH$_4$ fraction (%) | Pressure (mm Hg) | | | $L$ peak | $K$ peak | | | |
| Ar 1 | 6.5 | 695 | YCT 3 | 1124 | 0.346 | 0.386 | 129.1 | 104.0 | 0.870 |
| Ar 2 | 8.1 | 660 | YCT 5 | 1131 | 0.354 | 0.382 | 136.7 | 106.7 | 0.843 |
| Ar 3 | 8.5 | 710 | YCT 1 | 1164 | 0.347 | 0.388 | 151.7 | 138.0 | 0.882 |
| Ar 3-2 | 10.7 | 675 | YCT 8 | 1174 | 0.339 | 0.366 | 152.7 | 66.6 | 0.809 |
| Ar 4 | 7.4 | 645 | YCT 4 | 1138 | 0.353 | 0.377 | 166.8 | 147.8 | 0.830 |
| Ar 5 | 7.3 | 740 | YCT 9 | 1174 | 0.341 | 0.393 | 181.7 | 141.4 | 0.874 |
| Ar 6 | 8.8 | 670 | YCT 2 | 1139 | 0.352 | 0.381 | 196.7 | 150.5 | 0.846 |
| Ar 7 | 8.0 | 675 | YCT11 | 1128 | 0.356 | 0.388 | 211.7 | 135.0 | 0.830 |
| Ar 8 | 8.2 | 635 | YCT 8 | 1106 | 0.342 | 0.361 | 226.7 | 150.4 | 0.839 |
| Ar 9 | 7.9 | 715 | YCT 3 | 1154 | 0.344 | 0.388 | 241.7 | 136.6 | 0.853 |
| Ar10 | 8.1 | 655 | YCT 5 | 1126 | 0.355 | 0.381 | 256.7 | 149.3 | 0.802 |

cut was made from 15 m before to 3 h after every event that saturates the energy scale.

The energy windows for event selection were our standard 2 FWHM width (98% acceptance) and the window width in rise-time was set at our standard 96% acceptance. All events inside these windows during the entire period of counting were then considered as candidate $^{71}$Ge events. Examples of the resultant spectra are shown in Fig. 5. It is evident that the energy and rise-time windows include the vast majority of $^{71}$Ge events.

## VII. MEASURED PRODUCTION RATE

The number of selected events is given for each run in Table VII in the combination of both $L$ and $K$ peaks, and in Tables VIII and IX in the $L$ and $K$ peaks separately. The times of occurrence of the candidate $^{71}$Ge events were analyzed with our standard maximum-likelihood program [20] to separate the $^{71}$Ge 11.4-d decay events from a constant rate background. This is the same program that we used to analyze the runs with the $^{51}$Cr source and use to analyze all solar neutrino data. The likelihood function is given in [1]; the only analysis changes from the function used in the Cr experiment were to switch to the $^{37}$Ar half-life, to set the reference time to the present value, to slightly change the fixed solar neutrino rate to conform to the current best fit for all Ga experiments of 68.1 SNU [21], and to set the carryover term to zero. The latter was done because second extractions were made before each source exposure so there was no $^{71}$Ge carryover from one extraction to the next. The number of selected events that are fit to $^{71}$Ge is given in the second column of Tables VII, VIII, and IX. The remaining candidate events are assigned to background. The best fit assigns more than 200 events to $^{71}$Ge and has a signal to background ratio of 1.2. For comparison, in our $^{51}$Cr experiment, we assigned 144 events to $^{71}$Ge and had a signal to background ratio of approximately 1.0.

Note that the first extraction Ar 1 had a strange anomaly: 18.3 events were assigned to $^{71}$Ge in the $L$ peak, whereas only 2.8 events were so assigned in the $K$ peak. This appears to be just a statistical fluctuation as the counting system was func-tioning normally and examination of the data reveals no candidate events in the vicinity of the predicted *K*-peak position.

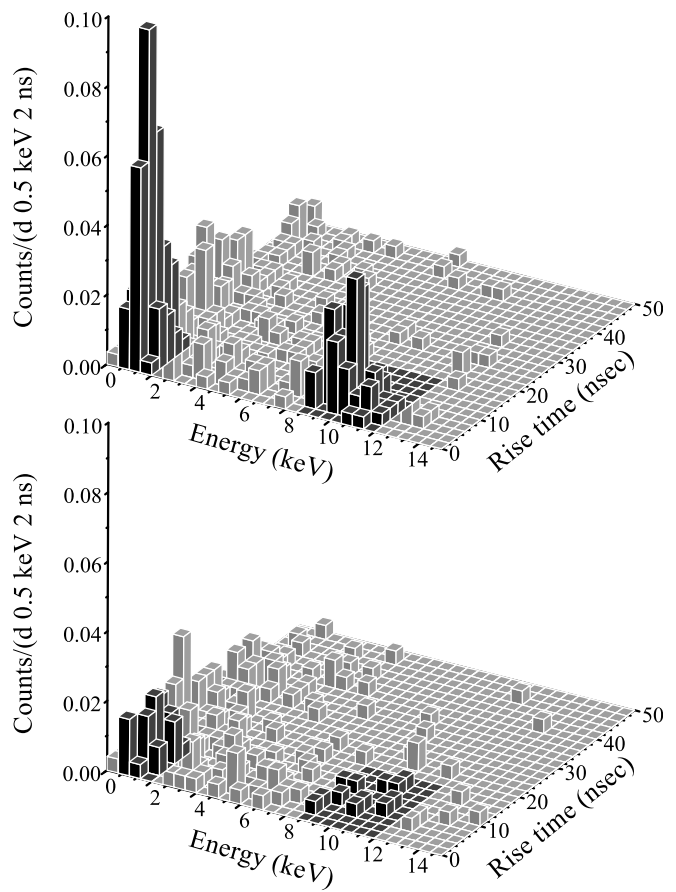

FIG. 5: Upper panel: energy vs rise-time histogram of all events after time cuts observed in all ten exposures during the first 30 days after extraction. The live time is 289.0 days and 472 events are shown. The expected location of the $^{71}$Ge $L$ and $K$ peaks based on the $^{55}$Fe calibrations is shown darkened. Lower panel: the same histogram for the 217 events that occurred during an equal live-time interval beginning at day 100 after extraction. The $^{71}$Ge has decayed away and is absent. The number of events outside the peaks is about the same in both panels as these are mainly due to background.

TABLE VII: Results of analysis of *L*- and *K*-peak events. All production rates are referred to the time of the start of the first exposure. The combined result excludes extraction Ar 3-2. The parameter $Nw^2$ measures the goodness of fit of the sequence of event times [22, 23]. The probability was inferred from $Nw^2$ by simulation.

| Extraction name | Number of candidate events | Number fit to $^{71}$Ge | Number of events assigned to $^{37}$Ar source production | Number of events assigned to Solar $\nu$ production | $^{71}$Ge production rate by $^{37}$Ar source (atoms/day) | $Nw^2$ | Probability (percent) |
|---|---|---|---|---|---|---|---|
| Ar 1   | 28  | 20.1  | 19.4  | 0.7  | $10.3^{+3.2}_{-2.8}$ | 0.065 | 60 |
| Ar 2   | 48  | 29.9  | 28.7  | 1.2  | $10.5^{+2.5}_{-2.2}$ | 0.048 | 73 |
| Ar 3   | 69  | 52.9  | 51.3  | 1.6  | $14.5^{+2.3}_{-2.1}$ | 0.110 | 35 |
| Ar 3-2 | 13  | 2.4   | 2.3   | 0.1  | $0.8^{+1.0}_{-0.8}$  | 0.273 | 7  |
| Ar 4   | 45  | 25.4  | 23.8  | 1.6  | $9.5^{+2.4}_{-2.2}$  | 0.142 | 13 |
| Ar 5   | 38  | 25.6  | 23.8  | 1.8  | $11.5^{+2.9}_{-2.6}$ | 0.108 | 29 |
| Ar 6   | 34  | 11.6  | 9.7   | 1.9  | $6.5^{+3.2}_{-2.7}$  | 0.042 | 81 |
| Ar 7   | 18  | 8.4   | 6.7   | 1.7  | $6.1^{+3.3}_{-2.7}$  | 0.079 | 43 |
| Ar 8   | 30  | 12.8  | 11.3  | 1.6  | $14.5^{+6.3}_{-5.4}$ | 0.051 | 72 |
| Ar 9   | 20  | 9.3   | 7.6   | 1.7  | $12.5^{+6.6}_{-5.5}$ | 0.066 | 82 |
| Ar10   | 39  | 7.4   | 5.8   | 1.6  | $13.6^{+9.2}_{-7.4}$ | 0.151 | 18 |
| Combined | 369 | 204.1 | 188.8 | 15.2 | $11.0^{+1.0}_{-0.9}$ | 0.048 | 77 |

TABLE VIII: Results of analysis of *L*-peak events. The production rate for each exposure is referred to its starting time. The production rate for the combined result is referred to the start of the first exposure. See caption for Table VII for further explanation.

| Extraction name | Number of candidate events | Number fit to $^{71}$Ge | Number of events assigned to $^{37}$Ar source production | Number of events assigned to Solar $\nu$ production | $^{71}$Ge production rate by $^{37}$Ar source (atoms/day) | $Nw^2$ | Probability (percent) |
|---|---|---|---|---|---|---|---|
| Ar 1   | 20  | 18.3  | 18.0  | 0.3 | $20.2^{+4.7}_{-5.8}$ | 0.094 | 53 |
| Ar 2   | 23  | 14.6  | 14.1  | 0.6 | $9.3^{+3.3}_{-2.6}$  | 0.068 | 56 |
| Ar 3   | 37  | 29.0  | 28.3  | 0.8 | $12.6^{+2.9}_{-2.2}$ | 0.048 | 78 |
| Ar 3-2 | 10  | 0.1   | 0.0   | 0.1 | $0.0^{+1.1}_{-0.0}$  | 0.311 | 8  |
| Ar 4   | 28  | 16.8  | 16.1  | 0.8 | $7.3^{+2.5}_{-1.7}$  | 0.093 | 30 |
| Ar 5   | 21  | 11.8  | 11.0  | 0.8 | $4.6^{+2.2}_{-1.4}$  | 0.076 | 45 |
| Ar 6   | 24  | 9.2   | 8.3   | 0.9 | $3.5^{+2.1}_{-1.2}$  | 0.047 | 72 |
| Ar 7   | 10  | 3.1   | 2.3   | 0.8 | $1.0^{+1.4}_{-0.5}$  | 0.097 | 35 |
| Ar 8   | 17  | 3.1   | 2.3   | 0.8 | $1.0^{+1.6}_{-0.7}$  | 0.052 | 74 |
| Ar 9   | 12  | 5.4   | 4.7   | 0.8 | $2.0^{+1.7}_{-0.9}$  | 0.068 | 51 |
| Ar10   | 27  | 0.8   | 0.0   | 0.8 | $0.0^{+1.5}_{-0.0}$  | 0.103 | 48 |
| Combined | 219 | 114.6 | 107.4 | 7.3 | $13.2^{+1.6}_{-1.5}$ | 0.078 | 50 |

TABLE IX: Results of analysis of *K*-peak events. The production rate for each exposure is referred to its starting time. The production rate for the combined result is referred to the start of the first exposure. See caption for Table VII for further explanation.

| Extraction name | Number of candidate events | Number fit to $^{71}$Ge | Number of events assigned to $^{37}$Ar source production | Number of events assigned to Solar $\nu$ production | $^{71}$Ge production rate by $^{37}$Ar source (atoms/day) | $Nw^2$ | Probability (percent) |
|---|---|---|---|---|---|---|---|
| Ar 1   | 8   | 2.8   | 2.5   | 0.4 | $2.5^{+2.8}_{-2.0}$  | 0.040 | 88 |
| Ar 2   | 25  | 15.2  | 14.6  | 0.6 | $8.9^{+3.2}_{-2.4}$  | 0.092 | 33 |
| Ar 3   | 32  | 24.1  | 23.2  | 0.9 | $9.2^{+2.4}_{-1.8}$  | 0.100 | 35 |
| Ar 3-2 | 3   | 2.6   | 2.5   | 0.1 | $1.2^{+0.6}_{-1.0}$  | 0.062 | 79 |
| Ar 4   | 17  | 8.6   | 7.8   | 0.8 | $3.3^{+1.9}_{-1.1}$  | 0.075 | 43 |
| Ar 5   | 17  | 13.8  | 12.9  | 1.0 | $4.7^{+1.8}_{-1.1}$  | 0.069 | 64 |
| Ar 6   | 10  | 3.7   | 2.7   | 1.0 | $1.1^{+1.4}_{-0.4}$  | 0.081 | 41 |
| Ar 7   | 8   | 5.1   | 4.3   | 0.9 | $1.7^{+1.4}_{-0.6}$  | 0.031 | 95 |
| Ar 8   | 13  | 9.2   | 8.4   | 0.8 | $3.5^{+1.7}_{-1.1}$  | 0.116 | 27 |
| Ar 9   | 8   | 4.2   | 3.3   | 0.9 | $1.3^{+1.3}_{-0.5}$  | 0.127 | 22 |
| Ar10   | 12  | 5.7   | 4.9   | 0.8 | $2.0^{+1.6}_{-0.7}$  | 0.085 | 39 |
| Combined | 150 | 91.2 | 83.2 | 8.0 | $9.3^{+1.3}_{-1.2}$  | 0.092 | 34 |

As presented in [18], the counting efficiency is calculated based on the measured volume efficiency, the fraction of counting gas that is GeH$_4$, and the counter pressure. The latter two parameters are given in Table VI. All counters used in these runs have had their efficiency directly measured with either $^{71}$Ge, $^{69}$Ge, or $^{37}$Ar. Some counters were calibrated with more than one of these isotopes.

The derived production rate of $^{71}$Ge from the source is in column 6 of Tables VII, VIII, and IX. For all runs combined the best fit rate is $11.0^{+1.0}_{-0.9}$ atoms of $^{71}$Ge produced per day by the source at the reference time. The stated uncertainty is purely statistical and is given with 68% confidence. Energy weight factors were used in this analysis in the same way as we analyze the solar neutrino data. Their effect on the overall result is, however, quite small: without weights the rate decreases by 1.1%.

Some of the systematic uncertainties that enter this result have been given in the foregoing. For most of the other systematic effects we adopt the same percentage values as we used for the $^{51}$Cr experiment. The results are given in Table X.

TABLE X: Summary of the contributions to the systematic uncertainty in the measured neutrino capture rate. Unless otherwise stated, all uncertainties are symmetric. The total is taken to be the quadratic sum of the individual contributions. For comparison, the statistical uncertainty in the result of the $^{37}$Ar experiment is $^{+9.0}_{-8.6}$%.

| Origin of uncertainty | Uncertainty (%) |
|---|---|
| Chemical extraction efficiency | |
|     Mass of added Ge carrier | 2.1 |
|     Amount of Ge extracted | 3.5 |
|     Carrier carryover | 0.5 |
|     Mass of gallium | 0.5 |
|     Chemical extraction subtotal | 4.1 |
| Counting efficiency | |
|     Calculated efficiency | |
|         Volume efficiency | 0.5 |
|         Peak efficiency | 2.5 |
|         Simulations to adjust for counter filling | 1.7 |
|     Calibration statistics | |
|         Centroid | 0.1 |
|         Resolution | 0.3 |
|         Rise time cut | 0.6 |
|     Gain variations | +0.5 |
|     Counting efficiency subtotal | +3.2, −3.1 |
| Residual radon after time cuts | −1.7 |
| Solar neutrino background | 0.4 |
| $^{71}$Ge carryover | 0.0 |
| Total systematic uncertainty | +5.2, −5.4 |

The quadratic combination of all these systematic uncertainties is $^{+5.2}_{-5.4}$%. The measured production rate in the $K$ and $L$ peaks, including both statistical and systematic uncertainties, is thus

$$p(\text{measured}) = 11.0^{+1.0}_{-0.9} \text{ (stat) } \pm 0.6 \text{ (syst)} \quad (2)$$

atoms of $^{71}$Ge produced per day. This result is not sensitive to any one of the extractions; e.g., if the somewhat anomalous extraction Ar 1 is deleted from the data set, the combined result increases by only 0.7%.

## VIII. PREDICTED PRODUCTION RATE

For a neutrino source of activity $A$, it follows from the definition of the cross section $\sigma$ that the capture rate $p$ of neutrinos in a material around the source can be written as the product

$$p = AD\langle L\rangle\sigma, \quad (3)$$

where $D = \rho N_0 f_I/M$ is the atomic density of the target isotope (see Table XI for the values and uncertainties of the constants that enter $D$), and $\langle L\rangle$ is the average neutrino path length through the absorbing material, which in the case of a homogeneous source that emits isotropically is given by

$$\langle L\rangle = \frac{1}{4\pi V_S} \int_{\text{absorber}} dV_A \int_{\text{source}} \frac{dV_S}{r_{SA}^2}. \quad (4)$$

In this last equation $r_{SA}$ is the distance from point $S$ in the source to point $A$ in the absorber and the source and absorber volumes are $V_S$ and $V_A$, respectively.

The Ga-containing reactor in which the $^{37}$Ar source was placed was nearly cylindrical, with a dished bottom. Based on accurate measurements of the reactor shape, the path length $\langle L\rangle$ was determined by Monte Carlo integration over the source and absorber volumes to be $72.6 \pm 0.2$ cm. The accuracy of this integration was verified by checking its predictions for geometries that could be calculated analytically and by noting that the measured Ga mass contained in the reactor volume agreed with that predicted by the integration. The sensitivity of $\langle L\rangle$ to the reactor geometry, to the position of the source in the Ga, and to the spatial distribution of the source activity were all investigated by Monte Carlo integration, and the uncertainty given above includes these effects.

Based on the source activity of $409 \pm 2$ kCi, and combining the uncertainty terms in quadrature, the predicted production rate is thus

$$p(\text{predicted}) = 13.9^{+1.0}_{-0.4} \quad (5)$$

atoms of $^{71}$Ge produced per day.

## IX. SUMMARY

The measured production rate for each run is plotted in Fig. 6 and the predicted rate is also shown for comparison. The sequence of measurements fits together very well with $\chi^2/\text{DOF} = 8.6/9$ (probability of 48%), where the comparison is made to the combined best fit of 11.0 atoms/d.

The ratio of measured to predicted production rates is

$$\frac{p(\text{measured})}{p(\text{predicted})} = \frac{11.0^{+1.0}_{-0.9} \text{ (stat) } \pm 0.6 \text{ (syst)}}{13.9^{+1.0}_{-0.4}} = 0.79^{+0.09}_{-0.10}, \quad (6)$$

where the statistical and systematic uncertainties have been combined in quadrature. This result is nearly $2.5\sigma$ less than unity which has a probability of slightly more than 1%.

To check that this unexpectedly low result is not the consequence of some experimental problem, all aspects of the experiment were carefully scrutinized, including remeauring the mass of Ga, verifying the position of the source within the Ga, and checking the extraction efficiency, the source strength, the counting efficiency, and the functioning of the counting system. No significant problems were found.

TABLE XI: Values and uncertainties of the terms that enter the calculation of the predicted production rate. All uncertainties are symmetric except for the cross section.

| Term | Value | Uncertainty Magnitude | Uncertainty Percentage |
|---|---|---|---|
| Atomic density $D = \rho N_0 f_I / M$ | | | |
|     Ga density $\rho$ (g Ga/cm$^3$) [24] | 6.095 | 0.002 | 0.033 |
|     Avogadro's number $N_o$ ($10^{23}$ atoms Ga/mol) | 6.0221 | 0.0 | 0.0 |
|     $^{71}$Ga isotopic abundance $f_I$ (atoms $^{71}$Ga/100 atoms Ga)[25] | 39.8921 | 0.0062 | 0.016 |
|     Ga molecular weight $M$ (g Ga/mol) [25] | 69.72307 | 0.00013 | 0.0002 |
|     Atomic density $D$ ($10^{22}$ atoms $^{71}$Ga/cm$^3$) | 2.1001 | 0.0008 | 0.037 |
| Source activity at reference time $A$ ($10^{16}$ $^{37}$Ar decays/s) | 1.513 | 0.007 | 0.5 |
| Cross section $\sigma$ [$10^{-46}$ cm$^2$/($^{71}$Ga atom $^{37}$Ar decay)] [8] | 70.0 | +4.9, −2.1 | +7.0, −3.0 |
| Path length in Ga $\langle L \rangle$ (cm) | 72.6 | 0.2 | 0.28 |
| Predicted production rate ($^{71}$Ge atoms/d) | 13.9 | +1.0, −0.4 | +7.0, −3.0 |

## X. DISCUSSION

The major purposes in making the $^{37}$Ar source reported here were to develop the technology of source fabrication, to prove that a very intense source could be made, and to elaborate several techniques for source intensity measurement. These goals were achieved, and the source was further used to measure the response of the SAGE detector to $^{37}$Ar neutrinos.

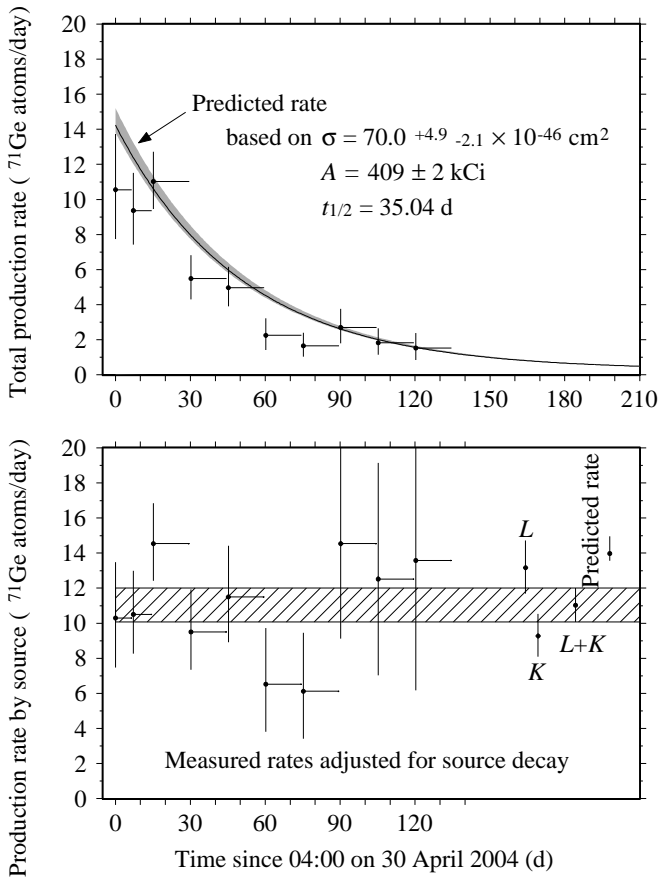

FIG. 6: Upper panel: comparison of measured total production rate for each extraction with predicted rate. Lower panel: measured rates from the $^{37}$Ar source extrapolated back to the start of the first extraction. The combined results for events in the the $L$ and $K$ peaks and for all events are shown separately at the right and compared to the predicted rate.

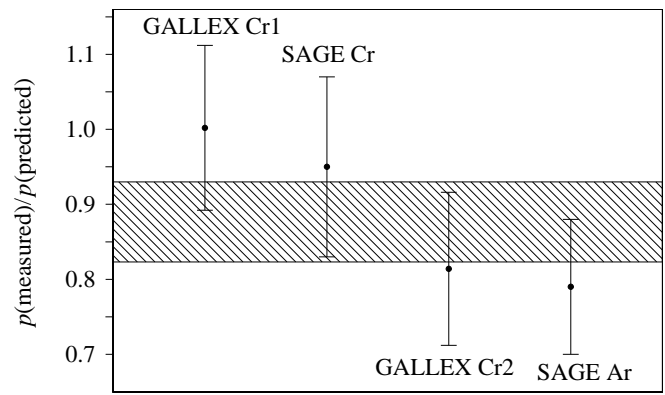

FIG. 7: Results of all neutrino source experiments with Ga. The hashed region is the weighted average of all four experiments.

The $^{37}$Ar source used in this experiment was made as a prototype for the production of a much more intense source. Based on the experience gained in making this source, the reactor engineers for BN-600 conclude that sources in the range of 2.0–2.5 MCi could be made if the Ca-containing modules were placed in the core of the reactor, rather than in the blanket zone, as was done here.

Since other experiments have given us great confidence in our knowledge of the various efficiencies in the SAGE detector, we do not consider this experiment to be a measurement of the entire throughput of SAGE. Rather, we believe this experiment should be considered in combination with the other source experiments with Ga and interpreted as a measurement of the cross section for the reaction $^{71}$Ga$(\nu_e, e^-)^{71}$Ge.

To this end, the results of the four neutrino source experiments with Ga given in Table I are shown graphically in Figure 7. The weighted average value of $R$, the ratio of measured to predicted $^{71}$Ge production rates, is $0.88 \pm 0.05$, more than two standard deviations less than unity. Although not statis-tically conclusive, the combination of these experiments suggests that the predicted rates may be overestimated.

Since 95% of the $^{71}$Ga neutrino absorption cross section simply depends on the $ft_{1/2}$ value for the transition from the ground state of $^{71}$Ge to the ground state of $^{71}$Ga, whose uncertainty is <0.5% [8], any error in the predicted rates must come from the contribution of the excited states. As discussed earlier, the Gamow-Teller strengths assigned to those transitions were deduced from $(p, n)$ cross sections, assuming a simple proportionality between $(p, n)$ and allowed weak interaction cross sections. Yet it is known phenomenologically that $(p, n)$ cross sections depend not only on the (weak interaction) Gamow-Teller amplitude, but also on a spin-tensor amplitude. Strong $(p, n)$ transitions require strong Gamow-Teller amplitudes, as the spin-tensor amplitude is generally a correction to the dominant Gamow-Teller term. In the case of a weak transition, however, it is possible that the spin-tensor amplitude dominates the $(p, n)$ cross section. There are several known examples of this, e.g., the $\ell$-forbidden M1 transition in $^{39}$K→$^{39}$Ca [26]. In this case the Gamow-Teller strength contributing to beta decay is very small, yet the $(p, n)$ cross section is appreciable and attributed to the presence of the spin-tensor interaction. In the case of $^{71}$Ga→$^{71}$Ge, the weak transitions to the excited states similarly could be due to the $\ell$-forbidden transition amplitude of the form $1f_{5/2}(n) \rightarrow 2p_{3/2}(p)$.

Thus, there is a theoretical uncertainty in the neutrino capture cross section, and it is quite possible that the Gamow-Teller strengths to the excited states are negligible, despite the nonzero $(p, n)$ cross sections [27]. As evidence for this hypothesis, we note that the weighted average of the four neutrino source experiments is $0.88 \pm 0.05$, reasonably consistent with $R = 0.95$, the value obtained if the excited state contribution were set to zero.

**Acknowledgments**

We thank Alexander Rumyantsev and Lev Ryabev (Federal Agency of Atomic Energy, Russia) and Valery Rubakov (Institute for Nuclear Research RAS, Russia) for their vigorous and continuous support for the $^{37}$Ar project. We are grateful to Wolfgang Hampel for helpful comments on a draft version of the manuscript and for providing revised results for the GALLEX source experiments. This work was partially funded by grant no. 05-02-17199 of the Russian Foundation for Basic Research as well as by grants from the USA, Japan, and Russia and carried out under the auspices of the International Science and Technology Center (project no. 1431). Additional funding came from the Program of Fundamental Research "Neutrino Studies" of the Russian Academy of Sciences.

[1] J. N. Abdurashitov, V. N. Gavrin, S. V. Girin, V. V. Gorbachev, T. V. Ibragimova, A. V. Kalikhov, N. G. Khairnasov, T. V. Knodel, V. N. Kornoukhov, I. N. Mirmov, A. A. Shikhin, E. P. Veretenkin, V. M. Vermul, V. E. Yants, G. T. Zatsepin, Yu. S. Khomyakov, A. V. Zvonarev, T. J. Bowles, J. S. Nico, W. A. Teasdale, D. L. Wark, M. L. Cherry, V. N. Karaulov, V. L. Levitin, V. I. Maev, P. I. Nazarenko, V. S. Shkol'nik, N. V. Skorikov, B. T. Cleveland, T. Daily, R. Davis, Jr. , K. Lande, C. K. Lee, P. S. Wildenhain, S. R. Elliott, and J. F. Wilkerson, Phys. Rev. C **59**, 2246 (1999) [arXiv: hep-ph/9803418].

[2] W. Hampel, G. Heusser, J. Kiko, T. Kirsten, M. Laubenstein, E. Pernicka, W. Rau, U. Rönn, C. Schlosser R. v. Ammon, K. H. Ebert, T. Fritsch, D. Heidt, E. Henrich, L. Stieglitz, F. Weirich, M. Balata, F. X. Hartmann, M. Sann, E. Bellotti, C. Cattadori, O. Cremonesi, N. Ferrari, E. Fiorini, L. Zanotti, M. Altmann, F. v. Feilitzsch, R. Mößbauer, G. Berthomieu, E. Schatzmann, I. Carmi, I. Dostrovsky, C. Bacci, P. Belli, R. Bernabei, S. d'Angelo, L. Paoluzi, A. Bevilacqua, M. Cribier, I. Gosset, J. Rich, M. Spiro, C. Tao, D. Vignaud, J. Boger, R. L.Hahn, J. K. Rowley, R. W. Stoenner, and J. Weneser, Phys. Lett. B **420**, 114 (1998).

[3] W. Hampel, private communication; E. Bellotti, *The Gallium Neutrino Observatory (GNO)*, presentation at TAUP 2003, 5–9 September 2003, Seattle, Washington, USA, http://www.int.washington.edu/talks/WorkShops/TAUP03/Parallel/. The values of $R$ in [2] ($1.01^{+0.12}_{-0.11}$ for GALLEX Cr1 and $0.84^{+0.12}_{-0.11}$ for GALLEX Cr2) have been revised due to changes in counter efficiencies, the solar neutrino subtraction, and the $^{222}$Rn cut inefficiency subtraction. (See also M. Altmann *et al.*, Phys. Lett. B **616**, 174 (2005) [arXiv: hep-ex/0504037].)

[4] M. Skalsey, Phys. Rev. C **39**, 2080 (1989).

[5] W. C. Haxton, Phys. Rev. C **38**, 2474 (1988).

[6] V. N. Gavrin, A. L. Kochetkov, V. N. Kornoukhov, A. A. Kosarev, and V. E. Yants, Institute for Nuclear Research of the Russian Academy of Sciences Report No. P-777 (1992).

[7] D. Krofcheck, E. Sugarbaker, J. Rapaport, D. Wang, J. N. Bahcall, R. C. Byrd, C. C. Foster, C. D. Goodman, I. J. Van Heerden, C. Gaarde, J. S. Larsen, D. J. Horen, and T. N. Taddeucci, Phys. Rev. C **55**, 1051 (1985).

[8] J. N. Bahcall, Phys. Rev. C **56**, 3391 (1997) [arXiv: hep-ph/9710491].

[9] F. B. Larkins, At. Data and Nucl. Data Tables **20**, 313 (1977).

[10] E. Huster and O. Krafft, Zeit. Naturforsch. A **24**, 285 (1969).

[11] J. P. Renier, H. Genz, K. W. D. Ledingham, and R. W. Fink, Phys. Rev. **166**, 935 (1968).

[12] C. E. Anderson, G.W. Wheeler, and W. W. Watson, Phys. Rev. **90**, 606 (1953).

[13] T. Lindquist and C-S Wu, Phys. Rev. **100**, 145 (1955); R. J. Glauber, P. C. Martin, T. Lindquist, and C-S Wu, Phys. Rev. **101**, 905 (1955).

[14] B. Saraf, Phys. Rev. **102**, 466 (1956).

[15] B. A. Zon, Yad. Phys. **13**, 963 (1971) [Sov. J. Nucl. Phys. **13**, 554 (1971)].

[16] W. Bambynek, H. Behrens, M. H. Chen, B. Crasemann, M. L. Fitzpatrick, K. W. D. Ledingham, H. Genz, M. Mutterer, and R. L. Intemann, Rev. Mod. Phys. **49**, 77 (1977).

[17] P. M. Endt, Nucl. Phys. A **521**, 1 (1990). See also *Half-Lives, Table of recommended values*, Laboratoire National Henri Becquerel Technical Note DIMRI/LNHB/01-2003 (2003), which gives $35.04 \pm 0.03$ d.

[18] J. N. Abdurashitov, V. N. Gavrin, S. V. Girin, V. V. Gorbachev, T. V. Ibragimova, A. V. Kalikhov, N. G. Khairnasov, T. V. Kn-

odel, I. N. Mirmov, A. A. Shikhin, E. P. Veretenkin, V. M. Vermul, V. E. Yants, G. T. Zatsepin, T. J. Bowles, W. A. Teasdale, D. L. Wark, M. L. Cherry, J. S. Nico, B. T. Cleveland, R. Davis, Jr., K. Lande, P. S. Wildenhain, S. R. Elliott, and J. F. Wilkerson, Phys. Rev. C **60**, 055801 (1999) [arXiv: astro-ph/9907113].

[19] S. Danshin, A, Kopylov, and V. Yants, Nucl. Instrum. Methods Phys. Res. A **349**, 466 (1994).

[20] B. T. Cleveland, Nucl. Instrum. Methods Phys. Res. **214**, 451 (1983).

[21] C. Cattadori, N. Ferrari, and L. Pandola, Nucl. Phys. B (Proc. Suppl.) **143**, 3 (2005).

[22] A. W. Marshall, Ann. Math. Stat. **29**, 307 (1958).

[23] B. T. Cleveland, Nucl. Instrum. Methods Phys. Res. A **416**, 405 (1998).

[24] H. Köster, F. Hensel, and E. U. Franck, Ber. Bunsen-Gs. **74**, 43 (1970).

[25] L. A. Machlan, J. W. Gramlich, L. J. Powell, and G. M. Lambert, J. Res. Natl. Bur. Stand. **91**, 323 (1986).

[26] J. W. Watson, W. Pairsuwan, B. D. Anderson, A. R. Baldwin, B. S. Flanders, R. Madey, R. J. McCarthy, B. A. Brown, B. H. Wildenthal, and C. C. Foster, Phys. Rev. Lett. **55**, 1369 (1985); B. A. Brown and B. H. Wildenthal, Atomic Data and Nuclear Data Tables **33**, 347 (1985).

[27] W. C. Haxton, Phys. Lett. B **431**, 110 (1998) [arXiv: nucl-th/9804011].